\documentclass[aps,prb,twocolumn,superscriptaddress,oneside,floatfix,amsmath,showpacs,amssymb]{revtex4-1}

\usepackage{graphicx}
\usepackage{dcolumn}
\usepackage{bm}
\usepackage{mathtools}
\usepackage[usenames]{color}
\usepackage{hyperref}
\usepackage[normalem]{ulem}
\begin{document}

\bibliographystyle{unsrt}

%\title{The influence of 1D topological states on the thermoelectric properties of Bi$_2$Te$_3$ in a framework of coupled helical Tomonaga-Luttinger liquids (TLLs)}
\title{Can 1D topological states explain the extraordinary thermoelectric properties of Bi$_2$Te$_3$ ? An exact analytical solution in the Tomonaga-Luttinger liquids (TLLs) framework}

\author{P.Chudzinski}
\affiliation{School of Mathematics and Physics, Queen's University Belfast}

\date{02.02.2020}

\begin{abstract}

Topological insulators are frequently also one of the best known thermoelectric materials. It has been recently discovered that in 3D topological insulators each skew dislocation can host a pair of 1D topological states – a helical TLL. We derive exact analytical formulas for thermoelectric Seebeck coefficient in TLL and investigate up to what extent one can ascribe the outstanding thermoelectric properties of Bi$_2$Te$_3$ to these 1D topological states. To this end we take a model of a dense dislocation network and find an analytic formula for an overlap between 1D (the TLL) and 3D electronic states.  %Any 1D state gives rise to huge Seebeck coefficient, a fact that is used in thermoelectricity enhancement by engineering of low dimensional nano-structures. One could then . However, in order to achieve a non-zero Seebeck coefficient, a mechanism that will induce curvature and backscattering of the 1D states is needed. One also needs a good description of 3D electronic states within a dense network of dislocations and a proof that a large overlap with the 1D states is possible. In this study we show how to overcome these obstacles and then derive an exact analytic formula for Seebeck coefficient within a framework of coupled TLLs. 
Our study is applicable to a weakly n-doped Bi$_2$Te$_3$ but also to a broader class of nano-structured materials with artificially created 1D systems. Furthermore, our results can be used at finite frequency settings e.g. to capture transport activated by photo-excitations. 

\end{abstract}

\pacs{
}

\maketitle

Thermoelectricity, on a fundamental level, gives us a valuable insight into interactions in a system. On the applications side it gives us hope to harvest electric energy from waste heat. The sensitivity to interactions manifest itself in the Mott relation, $S\sim (\partial_\omega\sigma(\omega))/\sigma(\omega)$, which links thermoelectric Seebeck coefficient $S$ and electric conductivity $\sigma$ in the adiabatic regime but also reveals how hard it is to describe (and improve) $S$ from the application viewpoint. In the single-particle picture when $\sigma(\omega)\sim LDOS(\omega)$, thanks to a singularity in the derivative of its density of states, the 1D metal could give rise to a huge Seebeck coefficient. This fact has been extensively used in thermoelectricity enhancement by engineering of low dimensional nano-structures\citep{Dresselhaus-1DTE} but the results of these attempts simply proved that unavoidably one needs a full description of electron-electron interactions in these systems. Bi$_2$Te$_3$, the first discovered strong 3D topological insulator\cite{Chen178} is also one of the best known thermoelectric materials at ambient conditions\cite{Snyder-rev}. When it was discovered\cite{1DTSon-disloc} that in such 3D topological insulator each dislocation will host a pair of counter-propagating 1D topological states – a helical Tomonaga-Luttinger liquid (TLL), there was momentarily an excitement that the outstanding thermoelectric properties of Bi$_2$Te$_3$ can be finally explained\cite{Sinova-weird1D}. However, several fundamental obstacles prohibit this simplistic picture. For the case of non-interacting fermions, or in a hydrodynamic liquid with an infinite cut-off, the thermoelectric transport coefficients are zero, which implies that a meaningful theory that predicts a substantial signal must include interactions and cut-off's in a non-perturbative manner. Furthermore, in order to achieve a non-zero Seebeck coefficient, a mechanism that will induce curvature and backscattering of the 1D states is needed. This is particularly difficult in helical TLL, where backscattering requires a spin flip. Incorporating a uniform spin-orbit coupling in the model, although it modifies the definition of Kramers-invariant variable, is insufficient as it does not allow for backscattering to occur\cite{Japaridze-inhomRash}.% [PRB 90, 075118].    

A series of recent results paved the way to overcome this stalemate. Most importantly it has been recently shown that a new type of Rashba-like spin-orbit coupling is present on a dislocation\citep{Hu-newSOdisloc}. The coupling has been shown to be momentum k-dependent, so it can be spatially inhomogeneous. If one now considers a periodic modulation of the Burgers vector (possibly accompanied by a local electric field) then the new Rashba term will be also periodically modulated in space. Such modulation can be induced by lattice distortion corresponding to a transverse optical (TO) phonon mode. This, and inelastic character of phonon processes\citep{Glazman-inel-backsc},\citep{Trauzettel-phonon-backsc}, is exactly the kind of term\citep{Japaridze-inhomRash} required to cause the desired spin-flip process and overcome the topological protection against backscattering\citep{Trauzettel-long-conditions}. This shall produce finite resistivities and in effect a finite Seebeck coefficient. 

Experimentally, it has been recently shown that each twin boundary in Bi$_2$Te$_3$ consists of a long chain of lattice dislocation\citep{Kim-disloc-netw-experim},\citep{Snyder-nanotwinn}. This proves that: i) dislocations are common in Bi$_2$Te$_3$ and material may host dense networks of dislocations that may turn the topological effect into a volume effect, ii) one can in principle control distances between dislocations by changing angle between crystal grains. Developments in experiment/numerics have quantified relation between dislocation and local strain field\citep{strain-disloc-chain}, while emerging field of dislons (quanta of dislocation motion) builds a link between the local strain and phonon-phonon interactions on dislocations\citep{Dresselhaus-dislons}. Finally, complex band structure of Bi$_2$Te$_3$ has been explored by advanced DFT methods (GW)\citep{CBM-GWabinitio} and confirmed by ARPES experiments\citep{ARPESandDFT},\cite{Zhang24}. A consensus has emerged that Bi$_2$Te$_3$ is an indirect narrow gap semiconductor with a bottom of a conduction band slightly off the Brillouin zone center\citep{CBM-GWabinitio}.

These achievements, in seemingly unrelated areas, have set up the following challenge for many body theory: to compute Seebeck effect in a helical TLL in a strong interaction, strong lattice anharmonicity regime accounting for the fact that the helical TLL states exist only in a finite-energy window (of order $\Lambda_0 \approx 0.25eV$) where they do not decay into energy-momentum matching bulk states. One also needs a good description of a dense network of dislocations to assess the feasibility of our theoretical proposal. In this paper we give an analytical formula for the strength of Seebeck coefficient due to network of helical 1D states at arbitrary frequency and temperature. The calculation proceeds in two stages: in the first stage we compute Seebeck coefficient of a single TLL state on a dislocation, while in the second stage we compute tunnelling probability between 3D states (with a given Fermi momentum $k_F^{3D}$) and the states localized on 1D dislocations' array. %T When the chemical potential is located within this window topological states should dominate the transport properties.  

%Intro (Thermoel, TI [Science 325, 178], 1D on disloc [Nat.Phys. 5, 298], thermoel 1D [PRB 47, 16631], past attempts [JPhysConfS 334, 012013]; non-uniform Rashba, new spin-orbit on a dislocation [PRL 121, 066401], TO phonons in Bi$_2$Te$_3$; networks of dislocations [Science 348, 6230])
%Our aim is to capture the physics of the following system. There is a dense network of 1D dislocations hosting each a helical TLL, these states are filled (chemical potential is close to the bottom of conducting band) and interact with each other via Coulomb interactions. There is also strong Coulomb interaction within each 1D topological state, as these electron-electron interactions are effectively unscreened (only these are present within the spectral gap). The 1D states do not hybridize with each other, but we permit a small phonon-assisted hybridization (tunnelling) of a tiny density of 3D carriers into 1D dislocation states.    

Our system consist of three parts: the 3D dilute electronic liquid from bulk conduction band with free fermion Hamiltonian $H_{3D}=\sum_k E_{DFT}^{(3D)}(\vec{k})c^{\dag}(\vec{k})c(\vec{k})$, the helical TLL $H_{TLL}^{1D}$(from the topological states on dislocation) and the TO phonons described by $H_{ph}$. The 1D theory is written in terms of collective modes, the density fields
$\phi_{\nu}(x)$ and canonically conjugate fields $\theta_{\nu}(x)$, with
$\Pi_{\nu}(x)=\partial_{x}\theta_{\nu}(x)$. These fields are directly related to respective fermionic densities $\partial_{x}\phi_{\nu}(x)=-\pi \rho_{\nu}(x)$. Then the TLL hamiltonian reads:
\begin{equation}\label{eq:ham-TLL-def}
    H_{TLL}^{1D}= \sum_{\nu} \int \frac{dx}{2\pi}
    \left[(v_{\nu}K_{\nu})(\pi \Pi_{\nu})^{2}+\left(\frac{v_{\nu}}{K_{\nu}}\right)(\partial_{x} \phi_{\nu})^{2}\right]
\end{equation}
where $v_{\nu},K_{\nu}$ are velocity and TLL parameter for all collective $\nu$-modes that constitutes our hydrodynamic liquid. For the helical TLL, because fermion spin and chirality are locked, there is only one mode so  we drop the $\nu$ index in the following. Electron-electron interactions that have purely forward character (due to the topological protection) are incorporated in $K=\sqrt{\frac{1-g}{1+g}}$ where $g=V_{Coul}(q\rightarrow 0)$. 

The TO phonon branch, considered here, in Bi$_2$Te$_3$ has the following properties: i) local boson density is well defined quantity as it corresponds to an amplitude of a local atomic oscillation within a unit cell; ii) these oscillations modify locally Burgers distortion and so the 'new-Rashba' term; iii) resonant bond character of the crystal lattice vibrations softens $\omega_0(q=0)$ \footnote{This also increases the strength of electron-phonon coupling which as usual scales as $\omega_0^{-3/2}$}; iv) when a TO phonon is being emitted/absorbed, causing electron backscattering from $+k_F$ to $-k_F$, then the dispersion $\omega(q_0 \approx 2 k_F)$ falls in the linear dispersion range (because of a typical available $k_F$ for our specific TLL). Hence the conditions to use the TLL theory for bosons are fulfilled and their Hamiltonian is: 
\begin{equation}\label{eq:ham-TLL-ph}
    H_{ph} = \int \frac{dx}{2\pi}
    \left[(v_{ph}K_{ph})(\pi \Pi_{ph})^{2}+\left(\frac{v_{ph}}{K_{ph}}\right)(\partial_{x} \phi_{ph})^{2}\right]
\end{equation}
where $\nabla \phi_{ph}$ describes the density of the \emph{local} $Bi-Te$ oscillations in a given, $x_i$-th, unit cell and the key advantage of using the TLL formalism is that any anharmonicity (which one expects to be large on the dislocation) can be captured by an appropriate choice of the TLL parameter $K_{ph}$. In the following we take $K_{ph}\approx 1$ which corresponds to hard-core bosons. The electron-phonon coupling reads:
\begin{equation}\label{eq:ham-el-ph}
H_R = \frac{g_R}{2\pi a} \int dx \partial\phi_{ph}(x)\exp(\imath \phi(x) )
\end{equation}
where we assumed, in accordance with the discussion in the introduction, that the phenomenon is proportional to the amplitude of distortion $\equiv$ density (spectral weight) of the TO phonons at a given point $n_{ph}=\partial\phi_{ph}(x)$. The amplitude of the process $g_R = V_R/\Lambda$ where $V_R$ has to be found from material specific \emph{ab-initio} calculation, in analogy with those in Ref.\citep{Hu-newSOdisloc}. We note that $V_R$ is of the same order as spin-orbit coupling in a given material and in Bi$_2$Te$_3$ while the spin-orbit coupling is also the underlying reason of band gap opening $\Delta_b$. Since $\Delta_b$  determines the $\Lambda$ UV-cut-off of the 1D hydrodynamics, the $g_R$ is expected to be not far from one. The filling of the system is incommensurate so we do not expect Eq.\ref{eq:ham-el-ph} to open 1D many-body gap at $E_F$, for a finite $\omega_0$ (non-adiabatic regime) the coupling is marginal\citep{Citro-adiab-nadiab}. This situation is ideally suited to employ memory function formalism\citep{Gotze-memory},\citep{SAR-fundament} where a single well-defined perturbation breaks the perfect conductivity of TLL. The charge/heat conductivities matrix is then expressed as:
\begin{equation}\label{eq:sigma}
   \hat{\sigma}(q,\omega;T)=\hat{\chi}(T)(-\imath\omega\hat{\chi}(T)+\hat{M}(q,\omega))\hat{\chi}(T)
\end{equation}
where $\hat{M}(q,\omega)$ and static susceptibilities $\hat{\chi}(T)$ are 2x2 matrix as there are two forces ($\nabla E_x$ and $\nabla T$) and two currents (electric and heat). The off-diagonal, thermoelectric conductivities, are equal by Onsager relation. The entries of $\hat{M}(q,\omega)$ are memory functions i.e. meromorphic functions each equal to a correlator of force-operators $M_{i,l}(q,\omega)=(\langle F_i F_l \rangle|_{q,\omega}-\langle F_i F_l \rangle|_{0,0})$ where the $\langle\rangle$ are computed for $H_{TLL}$. The force-operator $F_i(x,t)=[j_i(x,t),H_{tot}(x,t)]$, where $H_{tot}=H_{TLL}^{1D}+H_{ph}+H_R$, selects the term in the Hamiltonian that does not commute with the respective current. The Seebeck coefficient, defined in a stationary situation where both currents are zero (compensating forces), can be expressed as a ratio of the off-diagonal thermoelectric conductivity term $\sigma_{qe}$ and electric conductivity $\sigma_e$. Usually from the TLL description only the asymptotic behaviour at $T,\omega\rightarrow 0$ is extracted. However, for the purpose of this study, to be able to make a valid comparison with experiments, a full functional description valid at intermediate temperatures and frequencies is necessary. Explicit analytic expressions for $F(x,t)$ are given in the Supp.Mat. We used conformal field theory transformation to obtain the expressions valid at finite temperatures. In order to substitute these to Eq.\ref{eq:sigma} we need the Fourier transforms. Finding analytic form of the hyperbolic functions' transforms is the key outcome of our study. The $\int \exp(\imath (q' x +\omega t) \langle F_i(x,t) F_l(0,0) \rangle$ can be expressed as a convolution of electronic and phononic parts and for the uniform response $q'\rightarrow 0$ are a simple integral:
\begin{widetext}
\begin{multline}\label{eq:M-tot-eq}
M_{\text{eq}}(\omega',\text{T})\text{=}M^{(0)}T^{2 K+2+1-3} \int dq dw \Big[\Pi^{(ph)}\left(\frac{-v_{ph} q +w }{T},K_{ph}\right) \Pi^{(ph)}\left(\frac{v_{ph} q +w }{T},K_{ph}\right)\\
\left(\Gamma^{(el)}\left(\frac{w-q V_F}{T},K+1\right) \Pi^{(el)}\left(\frac{q V_F +w}{T},K\right)+\Gamma^{(el)}\left(\frac{q V_F +w}{T},K+1\right) \Pi^{(el)}\left(\frac{w-q V_F}{T},K\right)\right)\Big]
\end{multline}
\begin{multline}\label{eq:M-tot-qq}
M_{\text{qq}}(\omega',\text{T})\text{=}M^{(0q)}T^{2 K+2+2-3} \int dq dw \Big[\Pi^{(ph)}\left(\frac{-v_{ph} q +(w-\omega') }{T},K_{ph}\right) \Pi^{(ph)}\left(\frac{v_{ph} q +(w-\omega') }{T},K_{ph}\right)\\
\Big(\Gamma^{(el)}\left(\frac{q V_F +w}{T},K+1\right) \Gamma^{(el)}\left(\frac{w-q  V_F }{T},K+1\right)+\Gamma^{(el)}_{Q}\left(\frac{w-q V_F}{T},K\right) \Pi^{(el)}\left(\frac{q V_F+w}{T},K\right)+\\
\Gamma^{(el)}_{Q}\left(\frac{q V_F+w}{T},K\right) \Pi^{(el)}\left(\frac{w-q V_F}{T},K\right)+\Pi^{(el)}\left(\frac{q V_F +w}{T},K\right) \Pi^{(el)}\left(\frac{w-q V_F}{T},K+2\right)+\\
\Pi^{(el)}\left(\frac{q V_F+w}{T},K\right) \Pi^{(el)}\left(\frac{w-q V_F}{T},K+2\right)\Big)\Big]
\end{multline}
where the Fourier transforms of hyperbolic functions are $\Pi^{(el)}(z,Kr)= 2^K \frac{\beta\sqrt{v}}{2\pi} B_{(0,\exp(-\Lambda^{UV}_{tr} T))}\left(\frac{\pi  \text{Kr}+i z}{2 \pi },1-\text{Kr}\right)$, $\Pi^{(ph)}(z,Kr) =  2^1 \frac{\beta\sqrt{v_{ph}}}{2\pi}B_{(\exp(-\Lambda^{IR}_{tr} T),1.0)}\left(\frac{\pi\text{Kr}+i z}{2 \pi },1-\text{Kr}\right)$ and :
\begin{multline}\label{eq:M-el-ph}
\Gamma^{(el)}(z,Kr) = 2^{K+1} \frac{\beta\sqrt{v}}{2\pi} B_{(0,\exp(-\Lambda^{UV}_{tr} T))}\left(\frac{\pi  \text{Kr}+i z}{2 \pi },-\text{Kr}\right)+ 2^{K+1} \frac{\beta\sqrt{v}}{2\pi} B_{(0.,\exp(-\Lambda^{UV}_{tr} T))}\left(\frac{\pi  \text{Kr}+i z}{2 \pi }+1,-\text{Kr}\right)\\
\Gamma^{(el)}_{Q}(z,Kr) = 2^K \frac{\beta\sqrt{v}}{2\pi}\frac{\left(\frac{e}{2}\right)^{-K-2}}{(Kr-\imath z)\left(1+e^2\right)^2}~\times\\
F_1\left(\frac{1}{2} (\text{Kr}-i z);-2,\text{Kr}+2;\frac{1}{2} (\text{Kr}-i z+2);-\frac{1}{\exp(-2(\exp(-\Lambda^{UV}_{tr} T))},\frac{1}{\exp(-2(\exp(-\Lambda^{UV}_{tr} T))}\right)
\end{multline}
\end{widetext}
where $B_{x_1,x_2}(a,b)$ indicates a generalized incomplete \emph{Beta} function and $F_1(a;b;c;d;x_1,x_2)$ is Appell hypergeometric function. The temperature independent amplitudes $M^{(0)}$, $M^{(0q)}$ are given in Supp.Mat. This result generalizes well known single-mode TLL susceptibilities in two ways: i) it incorporates IR and UV transport cut-off's $-\Lambda^{UV,IR}_{tr}$  which in our problem are very close to the considered temperature/frequency range; ii) we have here Fourier transforms of the "vertex" functions\footnote{we introduce here notion of the "vertex" by using electron-phonon Ward identities definition, while in a pure electronic system that obeys Dzyaloshinskii-Larkin theorem the standard (non-transport) vertex functions are zero in TLL because RPA is exact and the chiral densities are conserved} $\sim \partial_x G_{TLL}(x,t)$ %The Appell function comes from the Fourier transform of $CotTanh^2(\xi_{\pm})$ ($\xi_{\pm}\equiv (x\pm V_{F} t)/T $) function, that enters the heat resistivity correlator, which has not been reported so far in the literature. We also generalized the previous TLL's Fourier transforms by taking generalized incomplete Beta functions instead of standard ones which captures the fact that our electronic system has an UV cut-off that is very close to the considered temperature/frequency range, while the phononic system has an IR cut-off due to finite $\omega_0$. In order to accommodate this fact in our formulas we imposed a cut-off in the integrals, details of calculations are in Supp. Mat..

\begin{figure}[ht]
	\centering
	\includegraphics[width=1.12\columnwidth]{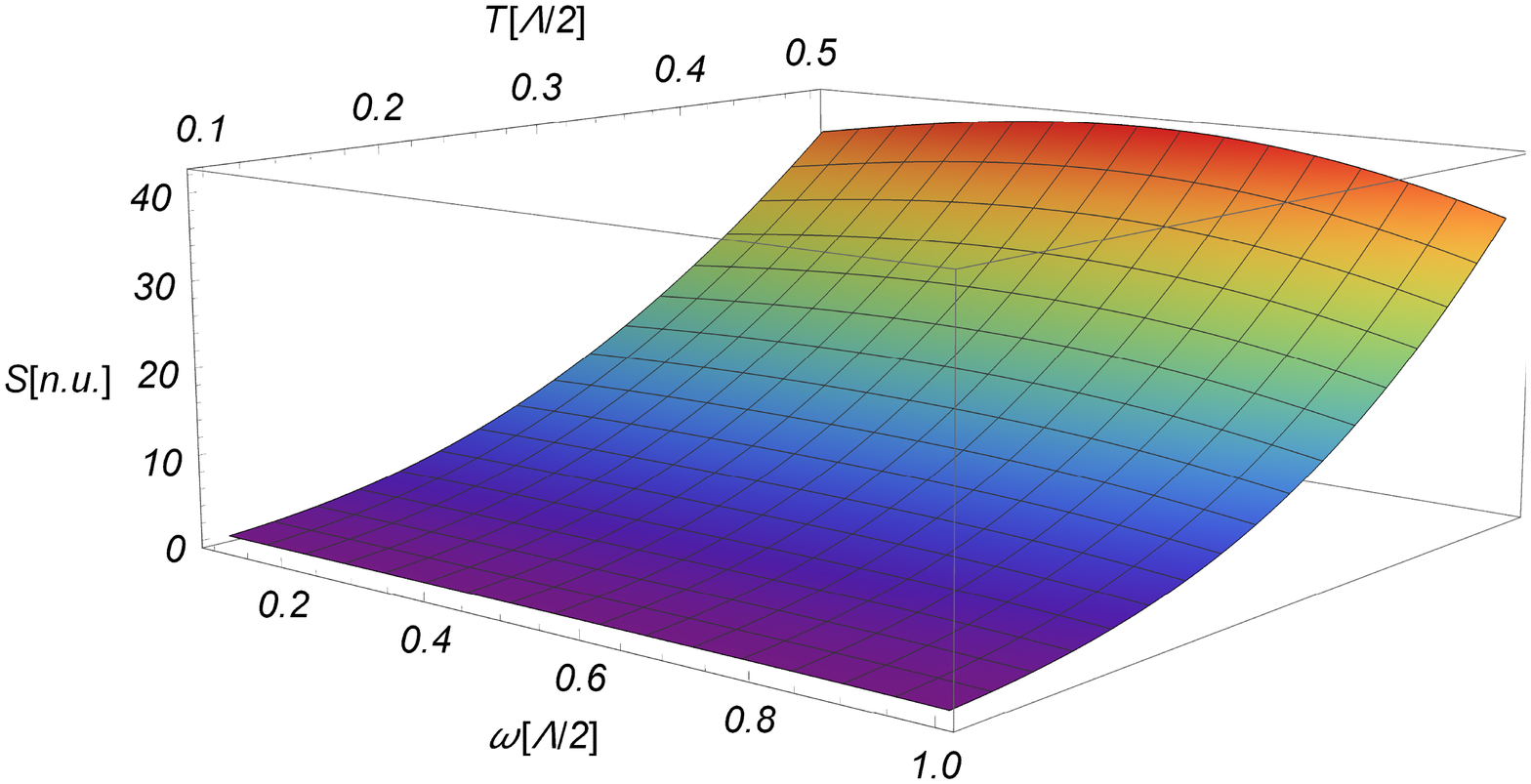}\\
	\includegraphics[width=1.12\columnwidth]{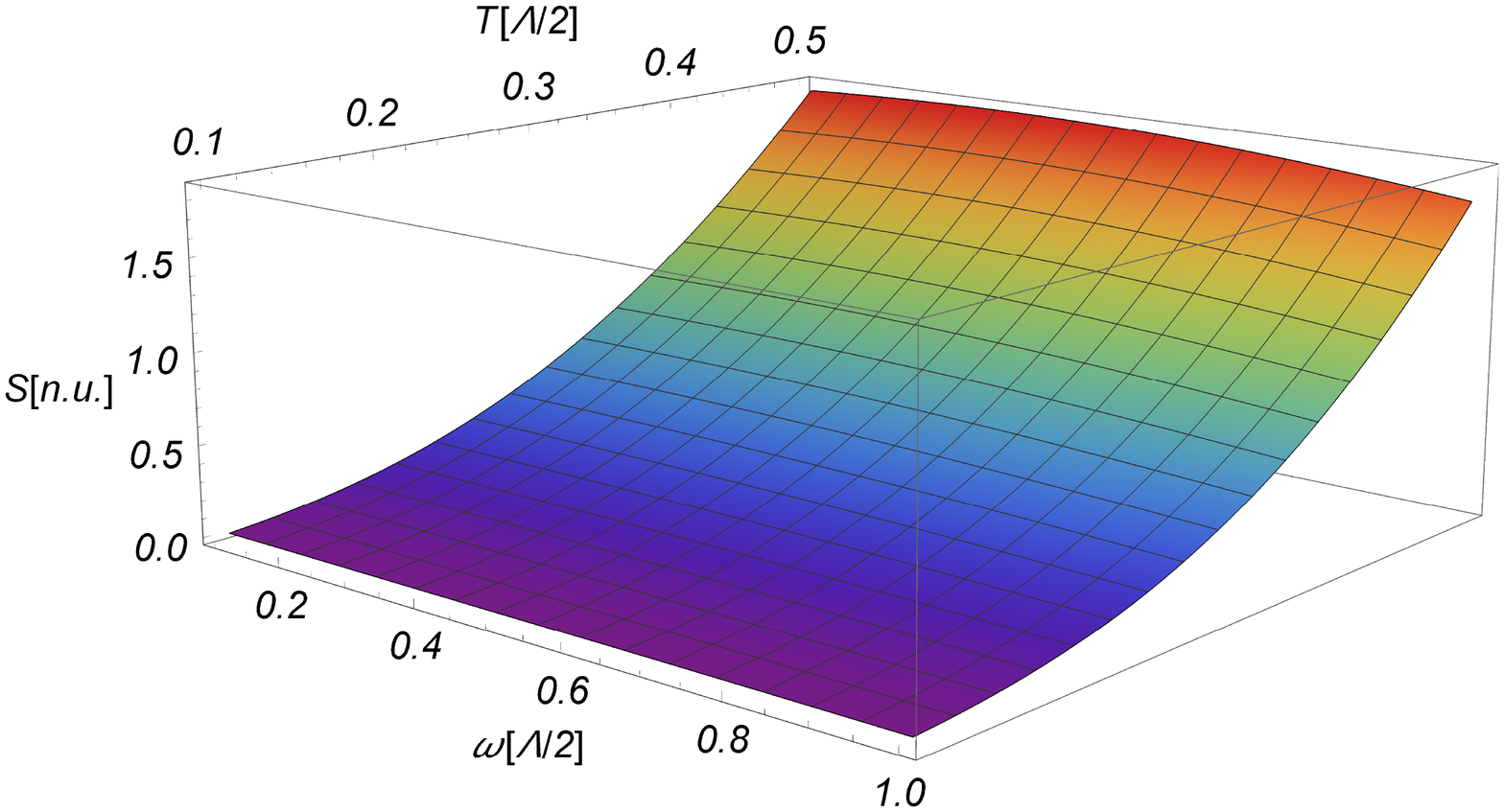}
	\caption{Seebeck coefficients calculated for 1D helical TLL subjected to Rashba type non-uniform spin-orbit coupling $V_R$. It is given in natural units and should be multiplied by $g_R^2$ where $g_R=V_R/\Lambda$. Here we compare the strength of thermoelectric effect for two different values of TLL parameter K=0.3 (top panel) and K=0.7 (bottom panel).}
	\label{fig:Seebeck-bare1D}
\end{figure}  

By substituting this into Eq.\ref{eq:sigma} we obtain a closed analytical formula for the Seebeck coefficient of TLL. We plot the result in Fig.\ref{fig:Seebeck-bare1D}. Since in our calculations we took $\hbar=e=k_B=1$, then the Seebeck coefficient is given in its natural units $1n.u.=28.6\frac{\mu V}{K} g_R^2$. Hence, if indeed $g_R\sim O(1)$, then the amplitude of the effect induced by dislocations can be quite substantial $~10^3\frac{\mu V}{K}$ which is on a par with the best thermoelectric materials.  On both panels we observe an increasing trend with temperature and apparently decreasing as a function of frequency. It should be noted that by increasing $v_{ph}$ (towards values that are unphysical for phonons in Bi$_2$Te$_3$ but may be realizable for other models) we are able to move the maximum towards finite frequencies. Remarkably, Seebeck effect increases (by an order of magnitude) when the TLL $K$ parameters decreases i.e. electron-electron interactions are stronger. Since the main factor that decreases $K$ are long-range Coulomb-type interactions then, assuming that the typical length of dislocation is large, it will be screening that determines the TLL parameter. This depends on the density of 3D carriers. We then predict that the Seebeck effect of dislocations will dramatically increase as the chemical potential approaches the conduction band minimum (CBM).

\begin{figure}[ht]
	\centering
	\includegraphics[width=1.12\columnwidth]{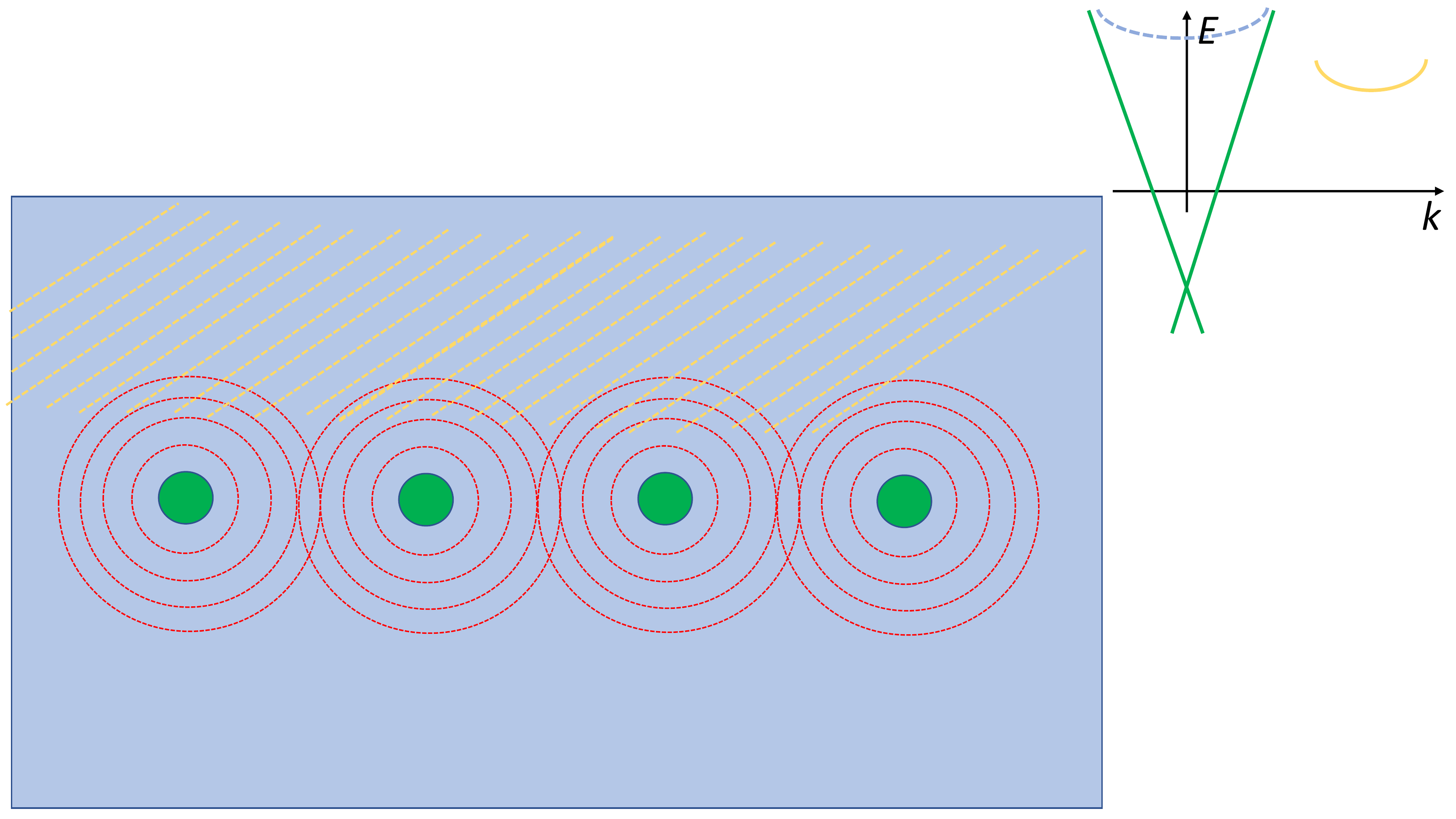}%\\
	\caption{Schematic model for the problem of tunneling into dislocation shown on the plane perpendicular to the dislocation axis. The incoming primary bulk wave $\Psi^{3D}_{1st}$ (yellow dashed lines) is scattered on a chain of dislocations (green) that is present on a crystalline twin-boundary. As a result of diffraction a secondary wave $\Psi^{3D}_{2nd}$ is produced (red dashed lines). Inset shows a sketch of Bi$_2$Te$_3$ band structure.}
	\label{fig:dislocation-model}
\end{figure}  

We can now move to the realistic setting where a network of dislocation is present in a weakly n-doped Bi$_2$Te$_3$, see Fig\ref{fig:dislocation-model}. Here we assume the following transport mechanism: an external electrode is connected to the 3D bulk states and these electrons, as they propagate through the sample, tunnel into the 1D dislocation states where the large thermoelectric coefficient can be harvested. We now incorporate the tunneling Hamiltonian into our description $H_{tun}=\sum_i t_{tun}(c^{\dag(3D)}(r_{\perp i})c^{(1D)}(r_{\perp i})+h.c.)$. From the standard procedure of constructing a second quantization Hamiltonian, the $t_{tun}$ $\equiv$ probability of tunnelling, is equal to an overlap between 3D and 1D wavefunctions on a given dislocation $\int dr_\perp\psi^{*3D}(r_\perp-r_{\perp i})\psi_{1D}(r_\perp-r_{\perp i})$. Taking $\psi_{1D}(r_\perp)=0$ if $r_\perp>R$ and constant inside this cylinder, implies that $t_{tun}$ is proportional to an amplitude of the 3D wave at the dislocation. The 3D electrons are primarily in the conduction band Bloch plane-wave states $\Psi^{3D}_{1st}(\vec{r})\sim\sum_{k\in k_{Fi}}\exp(\imath\vec{k}\vec{r})$, but while the Dirac 1D state is present around the $\Gamma$ point\footnote{if we take moderately anharmonic lattice then 1D electronic wave-function in $r_{\perp}$ plane is a combination of Hermite polynomials which can be approximated by a $Box(r_{\perp}/R_0)$ function and whose Fourier transform is $Sinc(q_{\perp})$, negative for $q_{\perp}\approx \pi/R_0$, so an overlap with any wave-packet at momentum of a fraction of BZ will be suppressed.}, the conduction band minimum (CBM) is located $\approx \pi/6 BZ$ away. The quasi-momentum conservation inhibits direct tunneling from the primary wave when $k_{\perp}^{(3D)}(CBM)\neq k_{\perp}^{(1D)}(\Gamma)$ because the two envelopes do not match. There are however also secondary $\psi^{3D}_{2nd}(r_{\perp})$, localized waves present due to multiple electron's wave scattering from the dislocations network.   

We take a plane wave coming from a pure crystal and compute multi-site diffraction pattern on cylindrical obstacles. The size of dislocations, distances between them and the Bloch wave-length are all comparable so one cannot consider a point-type wave-scattering but instead needs to use Fresnel diffraction of electronic waves. Such tunneling problem between 3D wave and 1D localized states has been solved in a closed analytical form in Ref.\cite{Chudzinski_2019}. The solution for the Fresnel diffraction on a circular aperture can be expressed as\cite{Hufford-Lommel-orig}:
\begin{eqnarray}\label{eq:Lomm-sol}
\psi_{2nd}(r)= \nonumber \\ 
\sum_{k_i}(\sin(N_{\bar{F}}^2(1+(r/R)^2)/2) + U_1(2 N_{\bar{F}}, 2 N_{\bar{F}} r/R))- \nonumber \\
 i (\cos(N_{\bar{F}}^2(1+(r/R)^2)/2) - U_2(2 N_{\bar{F}}, 2 N_{\bar{F}} r/R))
\end{eqnarray}

here $R$ is the radius of the dislocation, $r$ is a distance within the plane perpendicular to dislocation, $N_{\bar{F}}(k_i)=R^2/(\lambda\Lambda^{-1})$ is the Fresnel number, with $\lambda=\pi/k_i$ a wavelength of a Bloch-electron. To sum over $k_i$ we take two $k_F$'s at two sides of CBM along the direction of line of dislocations (presumably along the twin grain boundary). $U_{1,2}$ are Lommel functions of two arguments, $U_n (w,z) \approx \sum_{m=0}^{\infty} \left(\frac{w}{z}\right)^{n+2m}J_{n+2m}(z)$, here $J_{n+2m}(z)$ is a Bessel function of the first kind, they have a damped (weakly aperiodic) oscillatory behavior. Superposition of waves scattered on all dislocations gives the total amplitude $\Psi_{2nd}(r)=\sum_j\psi_{2nd}(r+jd)$ since each 1D system (with radius $R$) scatters electronic waves and hence becomes a source of an electronic wave $\psi_{2nd}(r)$.  %By superposition principle we add waves scattered by all 1D systems, this sum gives us $\psi_{f}(r>R)$. To be precise, we sum them up as a geometric series, but at the same time cut-off the long-distance component by including a finite escape length (imaginary self-energy of photo-electron) equal to two times unit cell size. For the core part of the \emph{final state}, $\psi_{f}(r<R)$, we assume a confinement within a parabolic quantum well which gives a Hermite polynomial solution for a wavefunction. The amplitude of this part is fixed by a boundary condition at $r=R$. The procedure has to be applied self-consistently: we write Hermite polynomials with given amplitudes which determines strength of each antenna at $r=R$, then we propagate the partial waves (Eq.\ref{eq:Lomm-sol}) using superposition rule Eq.\ref{eq:geom-final}, which in turn gives us a new boundary condition.
Since we are working at the very bottom of the conduction band we need to include Sommerfeld expansion for the temperature dependence of chemical potential\footnote{in addition there are DFT results for temperature dependence of the gap, that lead to similar effect}. As a result the $k_{F}$ does depend on temperature and in this indirect way temperature enters also Eq.\ref{eq:Lomm-sol}. 

\begin{figure}[ht]
  \centering 
  \includegraphics[width=0.95\columnwidth]{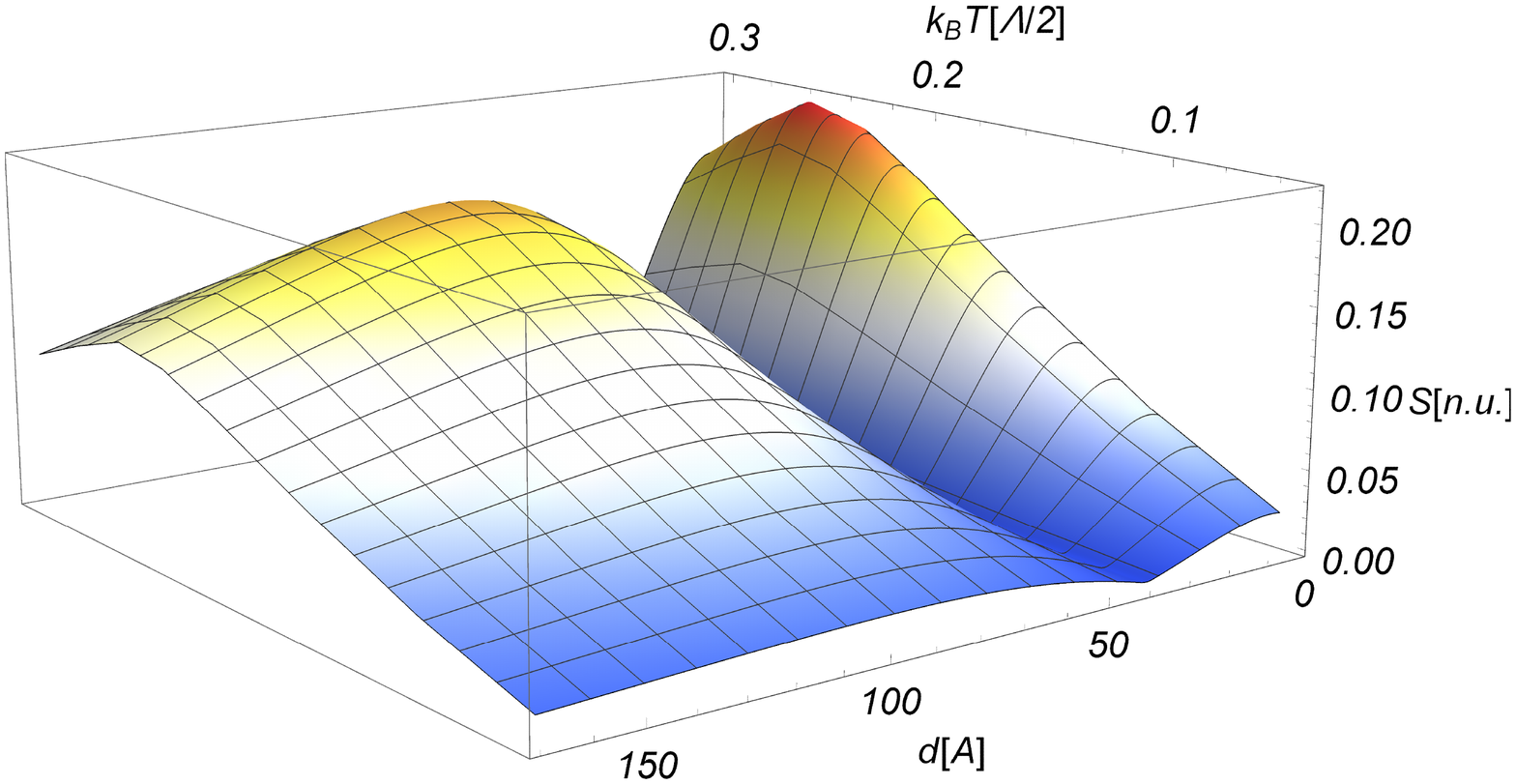}\\
  \includegraphics[width=0.95\columnwidth]{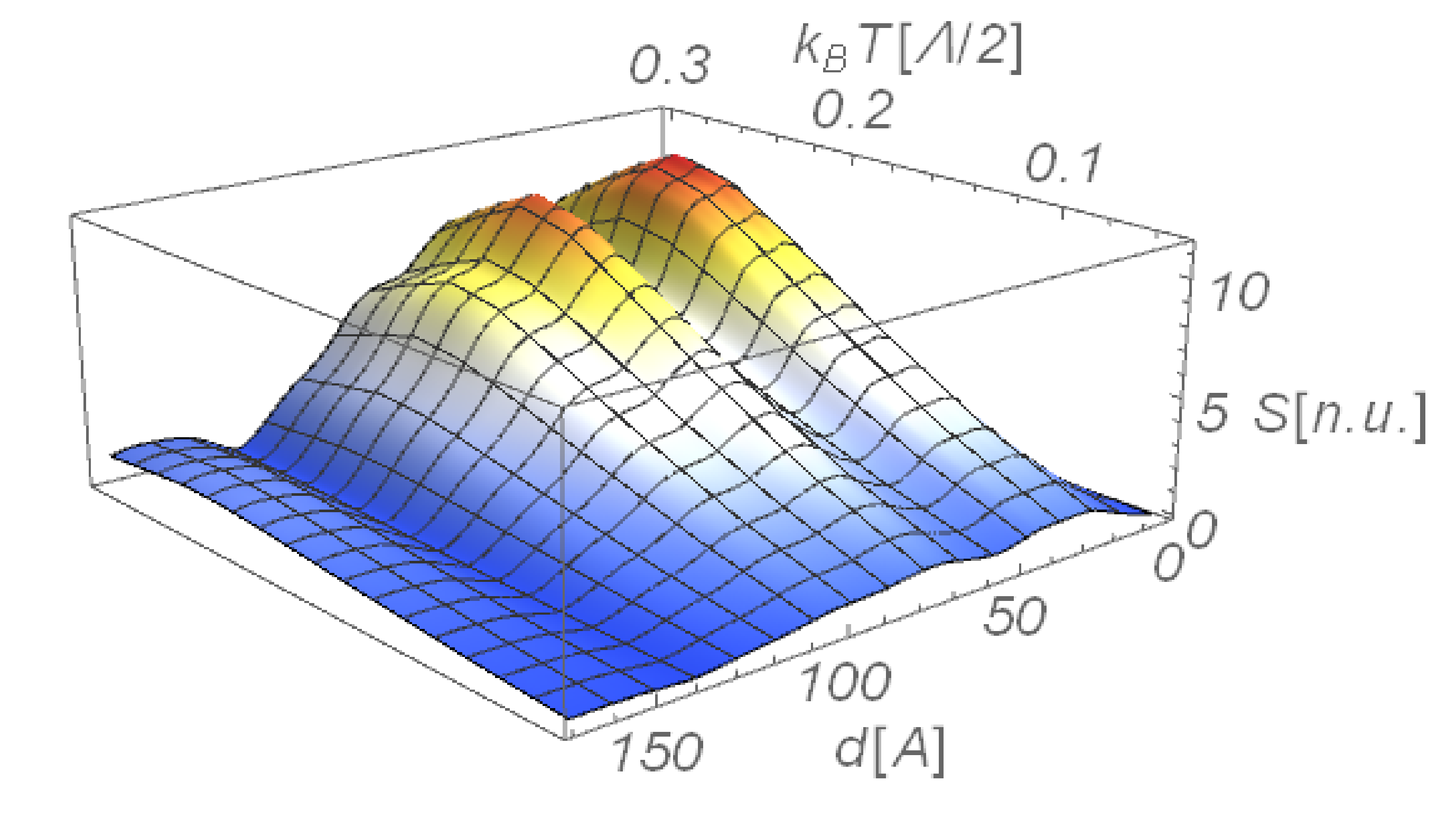}\\
	\caption{Seebeck coefficients, shown as a function of temperature and the inter-dislocation distance, calculated for a network of 1D dislocation embedded in a dilute 3D electron gas. In panel a) we take a realistic parameters for Bi$_2$Te$_3$ while in panel b) we tune the valley position to optimize the tunneling probability (see text).}
	\label{fig:LommSeeb}
\end{figure}

We can now combine Eq.\ref{eq:sigma} (with Eq.\ref{eq:M-tot-eq}-\ref{eq:M-el-ph}) and Eq.\ref{eq:Lomm-sol} to find the full temperature dependence of the Seebeck coefficient. The result of this calculation is shown in Fig.\ref{fig:LommSeeb}.  We see that instead of monotonically increasing $S(T)$ (as in Fig.\ref{fig:Seebeck-bare1D}) we have a broad maximum  which appears at around $0.25 \Lambda_0/2 \approx 350K$ which is in a reasonable agreement with experiment. Remarkably we also observe an evidence of the electronic waves interference phenomena -- the dependence on a distance between the dislocation is not monotonically increasing, as one would naively expect, but instead there is a complicated dependence with a well pronounced minimum at smaller distances. This is valuable information for the experimentalist: if an experiment is performed such that density of dislocation is varied in a controllable manner then one may find the system in a counter-intuitive regime where increasing the density reduces the Seebeck signal. We show here that this should not disprove the fact that there is a substantial contribution to Seebeck coefficient from dislocation. We also note that for the realistic parameters taken from experiment\citep{Kim-disloc-netw-experim} the amplitude of the effect is massively reduced, it becomes of order of $10\frac{\mu V}{K}$. This is in agreement with the amplitudes that were experimentally reported. The bottleneck of the transport mechanism is not an intrinsic property of Seebeck effect on a dislocation (which is by itself large) but instead it is suppressed by the extrinsic tunneling amplitude. This can be in principle modified by engineering methods. To explore this, we propose a hypothesis that the drastic suppression of the tunneling amplitude is related to two electronic waves $k_{CBM}\pm k_{F}^{(3D)}$ destructively interfering. To validate it we perform another calculation where one of the amplitudes is enlarged with respect to another. This is achievable, in our particular case, since $k_{CBM} - k_{F}^{(3D)} \approx 1/6 BZ$ so adding a superstructure with x6 periodicity can increase one of the amplitudes. An extra level of complexity that needs to be accounted here is due to an appearance of higher harmonics with momenta $k_n = k_{CBM} - k_{F}^{(3D)} + n\cdot 1/6 BZ$. The result of this calculation is shown on the bottom panel. We see that large values of Seebeck coefficient are recovered. This suggest the realistic way to harvest massive Seebeck coefficient from the dislocations.         

To conclude, firstly we have derived the exact analytic expressions for Fourier transforms of generalized correlation functions, Eq.\ref{eq:M-el-ph}, and for conductivity tensor, Eq.\ref{eq:M-tot-eq}-\ref{eq:M-tot-qq}, in TLL. The results of these show a dramatic departure from the prediction based on the Mott relation -- $\sigma(\omega)\sim \omega^a \implies S(\omega)\sim 1/\omega$, which shows that our formalism is able to capture profoundly non-adiabatic effects. Secondly, we have derived analytic formulas for tunneling between 3D electron liquid and a network of 1D TLLs. These are necessary ingredient in any transport device based on the 1D nano-structured states. Altogether the analytic solution allowed us to quantify the importance of topological states in thermoelectricity, provide an interpretation of recent experiments\citep{Kim-disloc-netw-experim}\citep{Snyder-nanotwinn} in Bi$_2$Te$_3$ and a pathway for future nano-structuring improvements that are available to be harvested.   

%These two results taken together make a showcase for the power of analytical methods and pave the way for their future applications in a broad class of transport phenomena. Thirdly, and perhaps more importantly from the point of view of applications, this study sets up a new perspective for large thermoelectricity present in Bi$_2$Te$_3$. We emphasize a very substantial role played by dislocations. We give a guidance for future experiments and DFT studies to investigate so far overlooked strength of electron-phonon coupling close to a dislocation and dynamic process of tunneling of polaron into "hot-electron". Finally, we provide an important insight for future applications showing an opportunity in harvesting thermoelectric heat at finite frequencies.

\bibliography{Bi2Te3TITE}

\end{document}